\documentclass[aps,twocolumn,showpacs,prl,floatfix]{revtex4}
\usepackage{graphicx}
\usepackage{epsf}
\usepackage{wrapfig}
\usepackage{epsfig}
\def\bk{{\mbox{\boldmath$k$}}}
\def\bq{{\mbox{\boldmath$q$}}}

\def\bp{{\mbox{\boldmath$p$}}}

 \def\br{{\mbox{\boldmath$r$}}}

   \def\bp{{\mbox{\boldmath$p$}}}

\def\br{{\mbox{\boldmath$r$}}}
\def\b0{{\mbox{\boldmath$0$}}}

\def\bk{{\mbox{\boldmath$k$}}}
\def\bq{{\mbox{\boldmath$q$}}}

\def\bp{{\mbox{\boldmath$p$}}}
\def\boldkappa{{\mbox{\boldmath$\Delta$}}}

\def\br{{\mbox{\boldmath$r$}}}
\def\b0{{\mbox{\boldmath$0$}}}

\newcommand{\ra}{\,\rangle}
\newcommand{\la}{\,\langle}

\def \b #1{ {\bf #1}}
\newcommand{\be}{\begin{eqnarray}}
\newcommand{\ee}{\end{eqnarray}}

\def \b #1{ {\bf #1}}

\def \b #1{ {\bf #1}}
     \font\tenbifull=cmmib10 scaled 1200 % bold math italic
     \font\tenbimed=cmmib9
     \font\tenbismall=cmmib7
       \def\bmit{\fam9 }
\mathchardef\bbkappa="7114
\mathchardef\bbrho="711A
\mathchardef\bbsigma="711B
\mathchardef\bbtau="711C
\mathchardef\bbvarrho="7125
\mathchardef\bbvarsigma="7126
\mathchardef\bbxi="7118
\def\boldkappa{{\bmit\bbkappa}}
\def\boldrho{{\bmit\bbrho}}

\begin{document}
\vskip 2mm \date{\today}\vskip 2mm
\title{Non-factorized calculation of the process
  $\bf ^3He(e,e'p)^2H$ at medium energies}
\author{C. Ciofi degli Atti}
\author{L.P. Kaptari}
\altaffiliation{On leave from  Bogoliubov Lab.
      Theor. Phys.,141980, JINR,  Dubna, Russia}
\address{Department of Physics, University of Perugia and
      Istituto Nazionale di Fisica Nucleare, Sezione di Perugia,
      Via A. Pascoli, I-06123, Italy}
 \vskip 2mm

\begin{abstract}
\vskip 5mm
The exclusive  process $^3He(e,e^\prime p)^2H$ has been analyzed
by a non-factorized and parameter-free approach  based upon
realistic few-body wave functions corresponding to the $AV18$
interaction
 and treating the rescattering
 of the struck nucleon
  within  a generalized eikonal approximation.
The results of our calculations, compared with recent JLab
experimental data,
 show that the left-right asymmetry  exhibits a clear dependence upon
 the  final state interaction demonstrating the breaking down of the
factorization approximation at  "negative" and large ($\gtrsim 300
\,MeV/c$)
 values of the missing momentum.
\end{abstract}
 \maketitle
Recent
experimental data obtained at JLab  on exclusive
electro-disintegration of $^3He$ \cite{jlab1} are at present the
object of intense theoretical activity (see
\cite{nashPR,Schiavilla,laget5} and References therein).
 In Refs. \cite{nashPR}   the 2-body, $^3He(e,e'p)^2H$, and 3-body, $^3He(e,e'p)np$,  break up
channels have been calculated
 within the following parameter-free approach:
i) initial state
  correlations (ISC)in the target nucleus $^3He$  have been taken care of  by using state-of-the-art
  few-body wave functions obtained ~\cite{pisa} by a variational solution of the Schr\"odinger
  equation containing realistic nucleon-nucleon interactions
   \cite{av18};
ii) final state interactions (FSI) have been treated by a
Generalized Eikonal Approximation (GEA)~\cite{mark}, which
represents an extended Glauber  approach (GA) based upon
 the evaluation of the relevant
  Feynman diagrams which describe  the rescattering of the
   struck nucleon in the final state, in analogy with the
  Feynman  diagrammatic approach  developed for the treatment
  of elastic hadron-nucleus scattering  \cite{gribov,bertocchi}.
In Ref.\cite{nashPR} the results of  calculations
 of the two-body break-up (2bbu) channel have been compared with  preliminary JLab data
 covering the region of "positive" (or "left") values of the missing momentum,
 i. e. the ones corresponding to $\phi=\pi$, where $\phi$ is  the   angle between
 the scattering and
 reaction planes.  Subsequent published experimental data  \cite{jlab1}
 covered, however,  also the region of "negative" (or "right")  values of the missing
 momentum, i.e. the ones
 corresponding to $\phi=0$,  which  have not been  considered in
  \cite{nashPR}. It is the  aim of this letter  to analyze the process in the
  entire experimental kinematical range and,  at the same time,
  to improve our theoretical approach.
As a matter of fact, our previous calculations \cite{nashPR} were
based upon the
 so called factorization
approximation (FA) in which the cross section is written as a
product of the free electron-nucleon cross sections and the nuclear
spectral function. Whereas the FA  is valid within
 the Plane Wave
Impulse Approximation (PWIA), it has been shown that it is
violated when FSI is  taken into account (see e.g.
\cite{UDIAS}-\cite{ udias1}). Within the FA, the $\phi$-dependence
  of the  cross section
   is only due to  the $\phi$-dependence
  of the  elementary
  cross section for  electron scattering off  a moving nucleon
  \cite{forest}. Such a dependence is in clear contradiction with the experimental data
  of Ref. \cite{jlab1} at  $p_m\gtrsim \, 0.35 \, \, GeV/c$, which is clear evidence
   of the breaking down of the FA. Performing non-factorized calculations within
relativistic and non-relativistic models for complex nuclei is nowadays a
common practice.
Non-factorized calculations for the process $^3He(e,e'p)^2H$ using
realistic non relativistic three-body wave functions
  have  only recently been   performed  \cite{laget5, Schiavilla},
  deriving the cross section
    in configuration
  space, with  different prescriptions for the non-relativistic reduction of the current
  operator. We have
  extended  our previous approach  \cite{nashPR} by relaxing   the FA
  and avoiding the problem of non-relativistic  reductions
  by directly performing  calculations in momentum space.
   We   do not consider, for the time being,
   Meson Exchange Currents (MEC), Isobar Configurations, and similar effects,
    which have been the object of intensive  theoretical studies;
     we fully concentrate on the effects of the FSI, treating  initial
and final state correlations,
   FSI and the current operator within a parameter-free
    approach.
By considering the general case of a target nucleus $A$, the
relevant quantities which characterize the 2bbu process
$A(e,e'p)(A-1)$ are the energy and momentum transfer $\nu$ and
$Q^2$, respectively,
   the
missing momentum\,\, ${\b p}_m =\bq - \bp_1$  (i.e.   the momentum
of the recoiling system  $A-1$), and  the\,\, {\it missing energy}
{ $E_m= \sqrt{P_{A-1}^2}+m_N -M_A \,\, = \nu - T_{A-1} -
T_{\bp_1}=|E_{A}| - |E_{A-1}| + E_{A-1}^*$. Here, $\bp_1$ is the
momentum of the detected proton and $E_A (E_{A-1})$ the (negative)
ground state energy of the target (recoiling) nucleus and
$E_{A-1}^*$ the intrinsic excitation energy of the latter
($E_{A-1}^* = 0$ in the 2bbu channel  we are going to consider).
The cross section of the process has form
\be
 \frac{d^6 \sigma}
  {d \Omega ' d {E'} ~ d^3{\b p}_m} = \sigma_{Mott} ~ \sum_i ~
V_i ~ W_{i}^A( \nu , Q^2, {\b p}_m, E_m)
 \label{cross}
 \ee
 where $i \equiv\{L,
T, TL, TT\}$,  $V_i$  are  kinematical factors~\cite{electron}
and  the nuclear structure functions $W_i^A$ result from proper
combinations of the polarization vector of the virtual photon,
$\varepsilon_{\lambda} ^{\mu}$,  and the hadronic tensor,
$W_{\mu\nu}^{A}$, the latter depending  upon the nuclear current
operators ${\hat J_\mu^A}(0)$. In the present  letter we describe
the two- and three-body ground states in terms of realistic wave
functions generated by modern two-body interactions~\cite{pisa},
and treat the final state interaction by a diagrammatic approach
of  the elastic rescattering of the struck nucleon with the
nucleons of the $A-1$ system. We consider the interaction of the
incoming virtual photon $\gamma^*$ with a bound nucleon (the
active nucleon) of low virtuality ($p^2\sim m_N^2$) in the
  quasi-elastic  kinematics i.e.  corresponding to $x\equiv Q^2/2m_N\nu\sim 1$. In
 quasi-elastic kinematics, the virtuality  of the struck nucleon after
    $\gamma^*$-absorption  is also rather low and,
    provided ${\b p}_1$  is sufficiently high, nucleon rescattering
    with the "spectator" $A-1$  can be   described to a large extent in terms of  multiple
    elastic scattering processes in the
    eikonal approximation. In co-ordinate representation the initial three-
    body wave function  is $\Phi_{1/2,{\cal
    M}_3}(\boldrho,\br)$, whereas the
wave function of the final state is
 \be
\Psi_f={\hat{\cal A}}\, S_\Delta^{FSI}(\boldrho,\br)\,e^{i {\bf
p}_1 {\bf r}_1}\, e^{i {\bf P}_D {\bf R}_D} \,\Phi_{1,{\cal
M}_2}(\br)\chi_{\frac{1}{2},s_f} \label{states}
 \ee
 where  $\boldrho$, $\br$ and ${\bf R}$ are usual  Jacobi
co-ordinates, ${\hat{\cal A}}$ denotes a proper antisymmetrization
operator, $\Phi_{1,{\cal M}_2}^*(\br)$ is the deuteron wave
function, and $\chi_{\frac{1}{2},s_f}$ the spin wave function of
the struck nucleon. The quantity
 \be
S_\Delta^{FSI}(\boldrho, \br)= 1 + T_{(1)}^{FSI}(\boldrho, \br)+
T_{(2)}^{FSI}(\boldrho, \br)
 \label{smatrix}
 \ee
is the GEA  S-matrix, which describes both the case of no FSI, as well as  single and double rescattering
 of the struck nucleon \cite{nashPR,mark}.
 For ease of presentation we will consider, from now-on, the
single rescattering contribution only, which has the form ${
T}_{(1)}^{FSI}(\boldrho,
\br)=-\sum\limits_{i=2}^3\theta(z_i-z_1){\rm e}^{i\Delta_z(z_i-z_1)} \Gamma (\b {b}_1-\b{b}_i)$,
where $\br_i \equiv(z_i, \b
b_i)$. It can be seen that, unlike the usual GA, the GEA S-matrix
gets also a contribution  from  a parallel momentum $\Delta_z$  of pure
nuclear origin and  depending upon the external
kinematics and
 the removal energy of the struck proton (within the "frozen approximation"  $\Delta_z=0$,  and
 the usual Glauber profile is recovered).
By assuming that the nuclear current operator is a sum of
nucleonic currents $\hat j_\mu(i)$, its momentum space
representation resulting from  Feynman diagrams within GEA,
can be written as follows
%the sum of two contributions
%$J_\mu^{3}=J_\mu^{3(1)}+J_\mu^{3(2)}$,
\begin{widetext}
\begin{equation}
 J_\mu^{3} =\sum_\lambda
 \int\frac{d {\bf p}}{(2\pi)^3} \frac{d\boldkappa}{(2\pi)^3}
S^{FSI}_\Delta (\bp,\boldkappa)
\la s_f|j_\mu(\boldkappa-\bp_m;\bq)|\lambda\ra
{\cal O} (\bp_m-\boldkappa,\bp; {\cal M}_3,{\cal M}_2,\lambda)=J_\mu^{3(PWIA)}+J_\mu^{3(1)}+J_\mu^{3(2)},
\label{ja}
\end{equation}
\end{widetext}
where $S_\Delta^{FSI}(\bp,\boldkappa) =\int d\br d\boldrho e^{-i\bp \br}
e^{i\boldkappa \boldrho} S_{\Delta}^{FSI}(\boldrho,\br)$, is the Fourier transform of the GEA
S-matrix (Eq. (\ref{smatrix})),
   ${\cal O}$ the nuclear overlap in
momentum space
\begin{widetext}
\be
\label{overl} {\cal O} (-\bk_1 = \bp_m -\boldkappa, \bp\,;
\,{\cal M}_3,{\cal M}_2,\lambda)= \int d\boldrho d\br e^{i(\bp_m -
\,\boldkappa)\, \boldrho} e^{i \bp\, \br}\Phi_{\frac{1}{2},{\cal
M}_3}(\boldrho,\br)\Phi_{1,{\cal M}_2}^*(\br)\chi_{\frac12
\lambda}^+.
\ee
\end{widetext}
and $\la s_f|j_\mu(\boldkappa-\bp_m;\bq)|\lambda\ra$ the nucleon current operator,
 $\lambda$ and $s_f$ being the spin projections of the struck
proton before and after $\gamma^*$ absorption. In absence of any FSI
 the momentum space S-matrix is $S_\Delta^{FSI}(\boldkappa ,
{\bf p})=(2\pi)^6\delta(\boldkappa)\delta({\bf p})$ and only the PWIA contribution
 $J_\mu^{3(PWIA)}$  survives in Eq. (\ref{ja}). When FSI is active,
 the contributions from single and double rescattering have to be taken into account; let
  us consider the former:  it results  from the single scattering momentum space term of
   $S_\Delta^{FSI}(\bp,\boldkappa)$, which has the following form
 \begin{widetext}
 \be
T_{(1)}^{FSI}(\bp,\boldkappa)= - \frac{(2\pi)^4}{k^*}
\frac{f_{NN}(\boldkappa_\perp)}{\boldkappa_{||}+\Delta_z-i\varepsilon}\left[\delta\left(
\bp-\frac{\boldkappa}{2}\right)+\delta\left(\bp+\frac{\boldkappa}{2}\right)\right]
\label{T1}
 \ee
 %\end{widetext}
 where $f_{NN}(\boldkappa_\perp)$, the Fourier transform of the profile function $\Gamma (\b b)$,
represents the
  elastic scattering amplitude of two nucleons with  center-of-mass
momentum $k^*$. Placing Eq. (\ref{T1}) in Eq. (\ref{ja}) one obtains the single scattering
contribution $J_\mu^{3(1)}$ to the nuclear current, {\it viz}
%\begin{widetext}
\be
&& \!\!\!\!\!\!\!\!\!\!\!\!\!\!\!\!
J_\mu^{3(1)}=\sum_\lambda  \int \frac{d\boldkappa}{(2\pi)^2 k^* }
\la s_f|j_\mu(\bk_1;\bq)|\lambda\ra
\frac{f_{NN}(\boldkappa_\perp)}{\boldkappa_{||} + \Delta_{||}-i\varepsilon}
%\times\nonumber\\&&
\left[
{\cal O} (-\bk_1,\frac{{\boldkappa}_\Delta}{2}; {\cal M}_3,{\cal M}_2,\lambda)
+{\cal O} (-\bk_1 ,-\frac{{\boldkappa}_\Delta}{2}; {\cal M}_3,{\cal M}_2,\lambda)
\right];
\label{single}
\ee
\end{widetext}
here   $\bk_1$ is the momentum of the proton before $\gamma^*$
absorption and   $\boldkappa =\bk_1+ \bq -\bp_1=\bk_1+\bp_m$   the
momentum transfer in the $NN$ rescattering.
 The longitudinal part of the
nucleon propagators has been computed using the relation
$[{\boldkappa_{||}+\Delta_z \pm i\varepsilon}]^{-1}=\mp i\pi\delta(\boldkappa_{||}+\Delta_z) +
PV[{\boldkappa_{||} +\Delta_z}]^{-1}$.
 Note  that in the eikonal approximation  the trajectory of
 the fast  nucleon is a straight line so that all
 "longitudinal" and "perpendicular" components are defined
 with respect  to this trajectory, which means that  the $z$ axis
 has  to be directed along the  momentum of the detected fast proton.
  Looking at the structure od Eq. (\ref{single}) it can be seen that due to the coupling of
the nucleonic current operator $\la
s_f|j_\mu(\bk_1;\bq)|\lambda\ra$ with the nuclear overlap integral
a factorized form for Eq. (\ref{single})  cannot be obtained; the
same holds for the double-scattering contribution and,
consequently, for the cross section.
 However, as shown in Ref.
\cite{nashPR}, if  the longitudinal part of the nucleonic current
can be disregarded, the factorization form can approximately be
recovered. Using the above formalism, and including the contribution from double rescattering $J_\mu^{3(2)}$,
the cross section (Eq.
(\ref{cross})) and the left-right asymmetry defined by
\be A_{TL}
=\frac{d\sigma(\phi=0^o)-d\sigma(\phi=180^o)}{d\sigma(\phi=0^o)+d\sigma(\phi=180^o)}.
\label{atl}
\ee have been calculated. Following  de Forest's
prescription \cite{forest}, the "CC1" form of the nucleon current
operator has been adopted; the elastic amplitude $f_{NN}$ has been
chosen in the  usual  form $f_{NN}(\boldkappa_\perp)=k^*
\frac{\sigma^{tot}(i+\alpha)}{4\pi}e^{-b^2\boldkappa_\perp^2/2}$,
where the slope parameter $b$, the total nucleon-nucleon cross
section $\sigma^{tot}$ and the ratio $\alpha$ of the real to the
imaginary parts  of the forward scattering amplitude, were taken
from world's experimental data.
 \begin{figure}[!hpt]
 \vskip -0.5cm
\includegraphics[height=8.0cm,width=8.5cm]{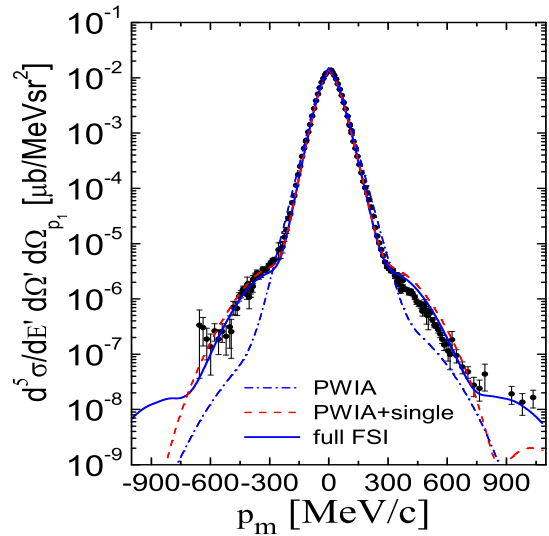}
\vskip -0.3cm \caption{The differential cross section for the
process $^3He(e,e^\prime p)^2H$ calculated within the
non-factorized approach compared with the PWIA result. {\it
Dot-dash}:
 PWIA; {\it dash}:
PWIA plus single rescattering ; {\it full}: PWIA plus single
and double rescattering. Experimental data from \cite{jlab1}.}
\label{Fig1}
\vskip -0.1cm
\end{figure}

The results of our calculations are shown
  in  Figs.\ref{Fig1}, \ref{Fig2} and \ref{Fig3}. In Fig.\ref{Fig1}
  the full non-factorized cross section
\begin{figure}[!hpt]\vskip -0.5cm
\includegraphics[height=8.0cm,width=8.5cm]{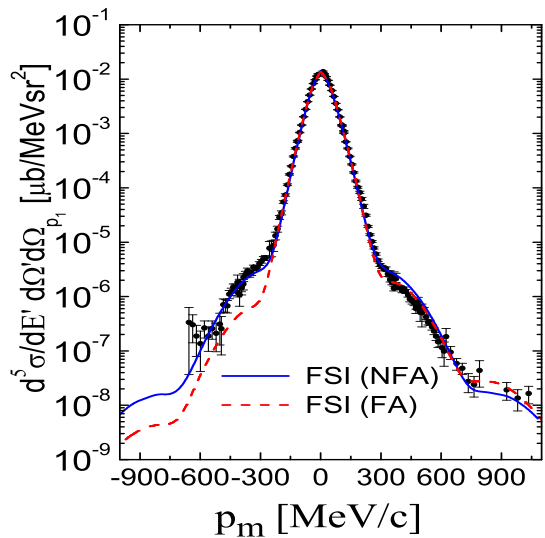}
\vskip -0.3cm\caption{The differential cross section for the
process $^3He(e,e^\prime p)^2H$ calculated taking into account FSI
within the non-factorized approach (FSI(NFA))  compared with the
results which include FSI within the factorized approximation
(FSI(FA)). Experimental data from \cite{jlab1}.} \label{Fig2}
\vskip -0.1cm
\end{figure}
is compared with the PWIA result;  Fig. \ref{Fig2} shows the
non-factorized and
  factorized cross sections;  in  Fig.\ref{Fig3}
   the asymmetry $A_{TL}$ is exhibited.  As in  previous calculations
   of ours,
   no approximations have
  been made in the evaluation
  of the single and double scattering contributions:
  intrinsic coordinates have been used and the energy dependence of the profile
  function  has been taken into account in the properly chosen CM system of the
  interacting pair.
In PWIA  the cross section is directly proportional to the two-body spectral function of  $^3He$.
 It  can be seen from Fig.\ref{Fig1}
 that  up to
 $|{\b p}_m|\sim 400\,\, MeV/c$,  the  PWIA and FSI results practically coincide  and, moreover,  fairly
 well agree with the experimental data; this  means  that
 the 2bbu process   $^3He(e,e'p)^2H$ does provide information on the two-body spectral function;
 on the contrary,
 at larger values of $|\bp_m|\gtrsim 400\,\, MeV/c$  the PWIA
 appreciably underestimates the experimental data. It is however very
 gratifying to see that   when the FSI is taken into account,  the disagreement
 is strongly reduced and an overall   good agreement
 between theoretical predictions and experimental data is obtained.
 It should be pointed out  that  at large values of the missing
 momentum the experimental data
 correspond to a  kinematics in which the deuteron momentum   is always
 almost perpendicular to the momentum of the final proton; in this case
  the effects of the  FSI are maximized; it  appears therefore that our approach
  is a  correct one in describing FSI effects.
The results presented in Fig.\ref{Fig2} clearly show that treating
FSI within the FA in the region  $\phi =0$ and $|\bp_m| \gtrsim
300 MeV/c$  is a poor approximation, unlike what happens at  $\phi
=\pi$. It should be pointed out that
 in spite of the good
agreement provided by the non-factorized  calculation,
  in some regions quantitative disagreements
with experimental data still persist. Particularly worth being
mentioned is the disagreement in the region around
 $|{\b p}_m|\simeq 0.6-0.65\,\, GeV/c$ at $\phi=0$.
 \begin{figure}[!hpt]\vskip -0.5cm
\includegraphics[height=7.5cm,width=8.5cm]{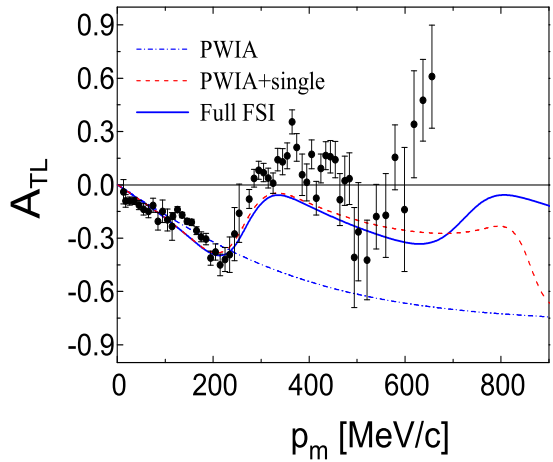}
\vskip -0.3cm\caption{The transverse-longitudinal asymmetry for the process
$^3He(e,e^\prime p)^2H$. {\it Dot-dash}: PWIA; {\it dash}: PWIA plus single rescattering FSI;
 {\it full}: PWIA
plus single and double rescattering FSI.
 Experimental data from ref. \cite{jlab1}}
 \label{Fig3}\vskip -0.1cm
\end{figure}
The origin of such a disagreement can better visualized by
analyzing the left-right asymmetry.
  It is well known that when  the explicit expressions of $V_i$ and $W_i^A$ are used in Eq.
  (\ref{atl})
 the
   numerator  is proportional to the transverse-longitudinal response $W_{TL}$, whereas the
 denominator does not contain $W_{TL}$ at all, which means that
 $A_{TL}$ is a measure of   the relevance of the transverse-longitudinal
 response relative to the other responses. In the
 $eN$ cross section the behavior of the asymmetry $A_{TL}$ is known
  to be a negative and  decreasing function of the missing momentum
  ~\cite{forest} and the same behaviour should be expected in  $eA$ scattering
  within the PWIA. The results presented in Fig. \ref{Fig3} clearly
 show that at $p_m \le 250 MeV/c$
  the PWIA result  is in  reasonable  agreement with the experimental
 data; however with increasing  $p_m$  the disagreement between
 the PWIA and  the experimental data
 appreciably   increases,  which seems to be a common features of all calculations so far performed  using microscopic three-body
 non-relativistic wave functions \cite{jlab1,Schiavilla};
 the origin of such a disagreement is at the moment  unclear (Meson Exchange currents
 considered in \cite{jlab1,Schiavilla}
  do not seem to solve this problem).
To sum up, we have calculated in momentum space the cross section of the
 processes  $^3He(e,e'p)^2H$, using realistic ground state two- and three-body
 wave functions and treating
 the FSI of the struck nucleon with the spectator nucleon pair
 within the  generalized eikonal  approximation.
 The method we have used is a very transparent
  and  parameter free  one: it
only requires the knowledge of
 the nuclear wave functions, since the FSI factor is fixed directly
 by  NN scattering data. At the same time, calculations are very involved
 mainly because of the complex structure of the wave function of Ref.~\cite{pisa},
which has to be firstly transformed to  momentum space and then
used in calculations of multidimensional integrals, including also
the computation of principal values. The main results of our
calculations can be summarized as follows: i) the violation of the
factorization approximation is appreciable at negative values
($\phi =0$) of $|\bp_m| \gtrsim 300 MeV/c$, whereas the
non-factorized and factorized results are in much better agreement
in the whole range of positive values ($\phi =\pi$)  of $|\bp_m|$;
ii) the left-right asymmetry can
 reasonably be reproduced
at low values of the missing momentum, but a substantial discrepancy between
theoretical calculations and experimental data
 remains to be explained at high values of $|\bp_m|$.

The authors are  indebted to  the Pisa Group   for making available
the variational three-body  wave functions.
Thanks are due  to S. Gilad, H. Morita,  E. Piasetzky, J. Ryckebusch,
 M. Sargsian, R. Schiavilla,  M. Strikman and J. Udias for
 many useful discussions.
L.P.K. is   indebted to  the University of Perugia and INFN,
Sezione di Perugia, for a grant and  warm hospitality. This
work was partially supported by the Italian Ministero
dell'Istruzione, Universit\`{a} e Ricerca (MIUR).

\end{document}